\documentclass{Interspeech}
\interspeechcameraready

\title{Towards explainable reference-free speech
intelligibility evaluation of people with pathological
speech}

\author[affiliation={1}]{Bence Mark}{Halpern}
\author[affiliation={2,3}]{Thomas}{Tienkamp}
\author[affiliation={2}]{Defne}{Abur}
\author[affiliation={1}]{Tomoki}{Toda}

\affiliation{}{Nagoya University}{Japan} 
\affiliation{}{Groningen University}{The Netherlands}
\affiliation{}{University Hospital Cologne}{Germany}

\email{halpernbence@gmail.com}

\usepackage{amsmath,graphicx}
\usepackage{cite}
\usepackage{amssymb,amsfonts}
\usepackage{algorithmic}
\usepackage{textcomp}
\usepackage{xcolor}
\usepackage{booktabs}
\usepackage{url}
\usepackage{enumitem}
\usepackage{tipa}
\usepackage{mdframed}
\usepackage{hyperref}

\begin{document}

\maketitle

\begin{abstract}
Objective assessment of speech that reflects meaningful changes in communication is crucial for clinical decision making and reproducible research. While existing objective assessments, particularly reference-based approaches, can capture intelligibility changes, they are often hindered by lack of explainability and the need for labor-intensive manual transcriptions. To address these issues, this work proposes the reference-free, explainable ASR Inconsistency Score. We evaluate this method on pathological speech in Dutch, Spanish and English, and compare its performance to a reference-based Word Error Rate (WER) baseline. Our results demonstrate that the ASR Inconsistency Score achieves a high correlation with expert perceptual ratings, with performance closely matching, and in one case exceeding, a standard reference-based Word Error Rate (WER) baseline.
\end{abstract}

\section{Introduction}

The reliable evaluation of speech plays a vital role in both therapeutic settings and clinical phonetic studies. For speech therapy, acoustic measures are important in guiding the monitoring of progress and the design of targeted interventions. Within clinical phonetics, these measures are crucial for ensuring research replicability and ascertaining the clinical impact of pathology on speech.

Speech assessment, in both clinical practice and research environments, can be divided into two approaches: perceptual (subjective) judgments and computational (objective) analysis. Perceptual evaluations involve listeners assessing various speech characteristics, such as intelligibility (defined here as the extent to which speech is comprehended). Nevertheless, such evaluations can be lengthy, require expert raters, and are susceptible to biases like prior listener exposure \cite{landa2014association} and level of professional expertise \cite{de1997test}. Conversely, computational evaluations use algorithms to examine speech signals and extract quantitative data, offering the prospect of automated, reliable, and efficient assessment of speech.

Among objective methods, reference-based approaches have been the most successful. A reference-based approach means that the evaluation requires a "clean" or "ideal" version of the speech (the reference) to compare against the speech being assessed. For instance, Pathological Short-Time Objective Intelligibility (P-STOI) and its extension, Extended STOI (E-STOI) \cite{janbakhshi2019pathological}, work by comparing the time-frequency representation of pathological input speech with a typical reference speech. Another approach, the Neural Acoustic Distance \cite{bartelds2022neural}, calculates the dissimilarity between self-supervised consensus representations of word segments from impaired and reference speech. Automatic Speech Recognition (ASR)-based methods use ASR systems to transcribe a pathological speech segment; the resulting transcription is then compared to the original text \cite{maier2009automatic,halpern2023automatic, tripathi2021automatic}. The error rate, such as Word Error Rate (WER), serves as an inverse measure of intelligibility. While not strictly reference-based in the signal processing sense, ASR methods require the correct, reference transcription.

However, acquiring a reference speech signal is difficult because it requires standardized read speech material (which may not accurately reflect real-world speech production \cite{keidser2020quest}) or the transcription of spontaneous speech (which is time-consuming and difficult, especially when speech is highly unintelligible).

To address this, reference-free methods can be used, such as supervised neural network-based approaches \cite{joshy2023dysarthria, 10094857, tripathi2021automatic, gupta2021residual, mallela2020voice}. However, these methods present their own challenges given that confounding variables are often present in the training data and prevent generalizability \cite{ozbolt2022things, schu2023using, liu24f_interspeech}. A further hurdle for their clinical adoption is the “black box” nature of these models, which lack the explainability needed to build clinical trust. To be truly viable in a clinical setting, systems should instead be composed of verifiable modules whose outputs are transparent and actionable. For example, phoneme log-likelihood is regarded as an interpretable proxy for articulatory precision by clinicians—even though the estimation process is a black box—because its output is clinically meaningful and verifiable \cite{liss2024operationalizing}.

As a step toward addressing these limitations, the current work proposes an explainable reference-free method called ASR Inconsistency Score. We show that the method can be used for pathological speech intelligibility estimation in three different languages.

In parallel to developing these measures, we also aimed to answer the following research questions (RQs):

\begin{enumerate}[label=\textbf{RQ\arabic*},noitemsep]
\item How does the proposed ASR Inconsistency Score perform compared to existing reference-based speech intelligibility methods and trivial confounder baselines for pathological speech? (Performance)
\item How does the choice of the language model used for estimating the ASR Inconsistency Score influence the final performance (Language model choice)?
\item How can we use the explanations provided by this model, and how accurate are the transcriptions generated by the model? (Explainability)
\end{enumerate}

\section{Proposed method}

Our method involves the use of two distinct ASR models.
The first model aims to transcribe the perceived, while the second aims to transcribe the intended message, thereby serving as a reference. Previous research found that ASR systems using robust language models tend to yield lower correlation results due to their deviation from a purely perceptual transcription \cite{halpern2023automatic, maier2009automatic}. However, to our knowledge, this characteristic has not previously been leveraged for the purpose of generating a reference transcription.

\subsection{Acoustic-Driven Transcription (\texorpdfstring{$W_{\text{greedy}}$}{Wgreedy})}

Both methods begin with a common acoustic model. Given an input acoustic feature sequence $X = (x_1, x_2, \ldots, x_T)$, a Connectionist Temporal Classification (CTC)-based acoustic model, $\mathcal{M}_{CTC}$, processes $X$ to yield a probability $P(y_t = k | X)$ for each symbol $k$ from the vocabulary $V'$ (which includes a 'blank' symbol $\epsilon$) at each time step $t$. The first transcription, representing the direct perceptual output ($W_{\text{greedy}}$), is generated via greedy decoding. This process selects the most probable symbol $\hat{y}_t$ at each time step, independent of context:
\begin{equation} \hat{y}_t = \text{arg max}_{k \in V'} P(y_t = k | X). \end{equation}

This creates a raw path $\hat{Y} = (\hat{y}_1, \ldots, \hat{y}_T)$. $W_{\text{greedy}}$ is then produced by applying a collapsing function $B$ to $\hat{Y}$, which merges consecutive identical symbols and removes all blank symbols, yielding $W_{\text{greedy}} = B(\hat{Y})$.

\subsection{Reference Generation Method 1: n-gram Beam-Search Decoding ($W_{\text{improved}}$)}

The first method for generating a reference transcription, $W_{\text{improved}}$, employs beam-search decoding integrated with a conventional $N$-gram language model ($P_{\text{LM}}$). The language model estimates the probability of a word sequence $W=(w_1, \ldots, w_M)$:

\begin{equation} P_{\text{LM}}(W) = \prod_{i=1}^{M} P(w_i | w_{i-N+1}, \ldots, w_{i-1}). \end{equation}

The beam search algorithm seeks to find a hypothesis sentence $W_h$ that maximizes a combined score of the acoustic and language models:

\begin{equation} \text{Score}(W_h) = \log P_{\text{CTC}}(W_h | X) + \alpha \log P_{\text{LM}}(W_h) + \beta \cdot \text{len}(W_h). \end{equation}

Here, $\alpha$ is the language model weight and $\beta$ is a word insertion bonus. The decoder explores a constrained set of hypotheses (determined by beam width $K$) and outputs the highest-scoring sentence as $W_{\text{improved}}$:

\begin{equation} W_{\text{improved}} = \text{arg max}_{W_h \in \mathcal{H}_K} \text{Score}(W_h). \end{equation}

\subsection{Reference Generation Method 2: Large Language Model (LLM) Correction ($W_{\text{LLM}}$)}

As a novel and possibly more powerful alternative, we tested the use of Large Language Models (LLMs) to generate the reference transcription. LLMs possess a far more sophisticated understanding of semantics, context, and grammar than N-gram models, possibly allowing them to better reconstruct the intended message.

In this approach, the initial greedy transcription, $W_{\text{greedy}}$, serves as an input to an LLM. Specifically, we prompted state-of-the-art models, including \texttt{gpt-3.5-turbo} and \texttt{gpt-4}, to correct the potentially erroneous text. The resulting corrected text is designated as the LLM-based reference, $W_{\text{LLM}}$. This process uses the LLM's world knowledge and linguistic capabilities to infer the speaker's original intent, producing an improved reference even from a highly distorted initial transcription.

\subsection{Final Score Calculation}

The final intelligibility score, $s_\text{score}$, is quantified by calculating the word error rate (WER) between the perceptual transcription ($W_{\text{greedy}}$) and the generated reference transcription ($W_{\text{improved}}$ or $W_{\text{LLM}}$). This error rate directly reflects the degree of inconsistency, where a higher score indicates a greater deviation of the acoustic signal from the intended message, and thus, lower intelligibility.

\section{Experimental settings}

\subsection{Evaluation datasets}

To evaluate our method, we selected two of our own datasets and two publicly available datasets. We selected datasets that had intelligibility annotations by experts at the sentence-level, not at the word-level or via diadochokinetic tasks. \textbf{None of the datasets below were used for training.}

\textbf{NKI-OC-VC}:
NKI-OC-VC \cite{halpern2023improving} contains the speech of 16 Dutch speakers (10 male, 6 female) who underwent a composite resection surgery or comparable treatment for mostly advanced tongue tumors. One participant's data was excluded from the analysis across all time points due to being recorded with a different, lower quality microphone, which compromised the recordings. This exclusion resulted in the 15 speakers whose data were ultimately analyzed.

Data collection occurred at up to three time points relative to their surgery: before surgery, within one month post-surgery, and approximately six months post-surgery. During three recording sessions, the participants did not complete the entire experiment. In total, the dataset features 26 speaker-time point combinations (e.g., speaker 1 pre-surgery).

Recordings took place during scheduled speech therapy sessions and were recorded with a Roland R-09HR field recorder at 44.1 kHz sampling frequency and 24-bit depth. This was later downsampled to 16 kHz and quantized to 16-bit. Participants were asked to read the Dutch text ``Jorinde en Joringel'' \cite{son01_eurospeech} consisting of 92 sentences during the recording session. One recording session (speaker/time point) lasted five minutes on average. The total duration of all speech recordings, across all speakers, was approximately 2.5 hours.

Speech severity ratings were provided by five SLPs using a five-point Likert scale with 5 denoting healthy speech, and 1 meaning severe speech problems. The interrater correlations between the intelligibility scores were excellent, so the scores were used for further analysis (\textit{ICC 2,k})=0.97.

\textbf{NKI-SpeechRT}: The original dataset \cite{speech_severity_evaluation_2024} comprised 55 speakers (45 male, 10 female) who had received concomitant chemoradiation therapy (CCRT) for head and neck cancer; of these, 47 were native Dutch speakers. One speaker was excluded from the analysis due to data processing issues. This resulted in the final cohort of 54 (44 male, 10 female) speakers included in the evaluation.

The dataset features recordings collected at a maximum of five distinct time points relative to the speakers' treatment. These time points include recordings pre-CCRT, 10 weeks post-CCRT, and 12 months post-CCRT. The specific timings of the remaining two potential recording time points are not detailed in the original dataset description. In total, the dataset features 138 speaker time points that represent the various combinations of these 54 speakers assessed across these available treatment time points.

Participants were asked to read a Dutch text
\cite{bomans2013vijvervrouw} which was cut into 23 segments for analysis. Recordings were made with a Sennheiser MD421 Dynamic Microphone and portable 24-bit digital wave recorder (Edirol Roland R-1). The speech samples were all downsampled to 16 kHz and quantized to 16-bit for later analysis. The total recorded data is approximately 4 hours.

Intelligibility ratings were provided by 14 recent Dutch SLP graduates. Participants rated the entire speech stimuli cut into three segments of approximately equal length. Intelligibility was rated on a 7-point Likert scale (1 = completely unintelligible, 7 = good) and showed excellent agreement \textit{ICC(2,k})=0.92. For a more detailed explanation of the experiment conditions, we refer the reader to Clapham et al's work \cite{clapham2012nki}.

\textbf{NeuroVoz}: The original dataset \cite{mendes2024neurovoz} contained Castilian Spanish speech recordings from 53 individuals with Parkinson's disease (PD; 33 male and 20 female) and 55 individuals without PD (28 male and 26 female). We used the data from 50 speakers with PD (30 male and 20 female) that had at least 13 overlapping sentences and corresponding intelligibility ratings.

Speech was recorded with an AKG C420 headset microphone connected to a preamplifier equipped with phantom power. The preamplifier was attached to a SoundBlaster Live sound card, which operates at a sampling rate of 44.1 kHz. This was later downsampled to 16 kHz sampling frequency and quantized to 16-bit.

Intelligibility ratings were provided by a single expert SLP with 15 years of experience. Intelligibility was graded on a scale of 1 to 5 for each utterance according to the original description. However, we could only identify 3 unique intelligibility levels, therefore we processed the scores on a 3-point scale for each utterance. We calculated a speaker-level score by taking the mean over the used utterances for the experiments.
With only one expert rater, no interrater correlation was calculated.

\textbf{TORGO}: The original TORGO \cite{rudzicz2012torgo} dataset contained Canadian English speech recordings from 8 individuals with dysarthria secondary to cerebral palsy and Amyotrophic Lateral Sclerosis (3 female, 5 male) and 7 typical speakers (3 female, 4 male). The selection of the 6 speakers and 55 sentences was based on the availability of overlapping transcriptions. Specifically, these 55 sentences were found to be common across the recordings of these 6 particular speakers (1 female, 5 male), and therefore, only these sentences from these speakers were included in the evaluation.

Among the various recording setups in the dataset, head-mounted microphone recordings were used (16 kHz), as this microphone provided a better selection of speakers. The dataset includes several different kinds of stimuli with orthographic descriptions provided. The provided annotations contained punctuation errors and typos that we have corrected to make the references identical.

A single clinician provided 9-point scale sentence intelligibility ratings according to the FDSA \cite{enderby1980frenchay}. As there was only one rater, no interrater correlation could be calculated.

\subsection{Implementation details}

\textbf{ASR systems.} To maintain a comparable ASR setting for each language, we use publicly available pre-trained HuggingFace models trained on the respective Common Voice resources \cite{ardilacommon}.\footnote{\url{jonatasgrosman/wav2vec2-large-xlsr-53-dutch,spanish,english}}.

\textbf{N-gram models}. For the beam-search decoding, we use the implementation provided by \texttt{pyctcdecode}. For comparability, each language model is used with the default $\alpha=0.5$ and $\beta=0.5$ hyperparameters. We have considered tuning these parameters in preliminary experiments, however, it did not bring any improvements.
The English and Dutch models are 5-grams trained on the respective language Wikipedia.\footnote{\url{BramVanroy/kenlm_wikipedia_en}}\footnote{\url{BramVanroy/kenlm_wikipedia_nl}} For the Spanish model, a different publicly available 5-gram is used.\footnote{\url{kensho/5gram-spanish-kenLM}}

\textbf{GPT models}.
We have used a self-written prompt as detailed in the frame below. We have tried optimizing the prompt using the official prompt corrector, however, it did not lead to improved results. As OpenAI API does not provide any interface for reproducible experiments, we decided to run the experiments 3 times with a temperature of 0 to account for the stochastic nature of the experiments. We report the mean and the 95\% confidence interval of the experiments, and perform a \texttt{t-test} to show if there is a significant difference in performance between the stochastic language models.

\begin{mdframed}
\textbf{Prompt:} The following is the output of an automatic speech recognition system for an utterance of a
speaker with speech pathology in [Dutch/English/Spanish]: [Sentence]

Please correct the sentence. Please put the corrected sentence within square brackets [like this].
If the sentence is already correct, repeat the sentence within square brackets.
\end{mdframed}

\subsection{Evaluation protocol}

To evaluate the performance of our models (\textbf{RQ1}) and the impact of possible confounders, we compare to the following measures:

\noindent \textbf{Speech Rate}: Calculated as the number of words in the ground-truth orthographic transcription divided by the total duration of the audio utterance (e.g., in words per minute).

\noindent \textbf{WADA-SNR}:
A reference-free Signal-to-Noise Ratio (SNR) estimation method which assumes that clean speech exhibits a Gamma distribution while additive noise is Gaussian \cite{kim08e_interspeech}.

\noindent \textbf{WER}: We compared our proposed system to the reference-based word error rate measure. The word error rate compares the transcription generated by the ASR system for the test audio against the ground-truth reference transcription. To evaluate the choice of the language model (\textbf{RQ2}), we will simply compare the performance of the respective models.

Finally, in order to show how one can explain the model decisions, we will show some example $W_{\text{greedy}}$ and $W_{\text{LLM}}$ pairs. To measure the accuracy of the explanations/reference (\textbf{RQ3}), we measured the WER between the $W_{\text{LLM}}$ and the ground truth reference ($W_{\text{LLM}}$ WER). A lower word error rate means the explanation is more correct. We also measure the WER between the $W_{\text{greedy}}$ and the ground truth reference ($W_{\text{LLM}}$ WER) for comparison. Finally, we also investigate how much the correctness of the corrected hypothesis depends on the intelligibility by reporting the correlations with the intelligibility scores.

\section{Results and Discussion}

\subsection{RQ1: Performance}

\begin{table}[htbp]
\centering
\caption{Pearson's correlation (r) of the speaker-level intelligibility scores with the ground truth perceptual intelligibility scores for the proposed and baseline systems. The best system in each column is highlighted in \textbf{bold}. The best reference-free system is \underline{underlined}. $n_{\text{spk}}$ indicates the number of unique speakers, while $n_{\text{spk-time}}$ means the number of unique speaker-time combinations. $n_{\text{sen}}$ indicates the number of sentences used for analysis per speaker. $^\dagger$ indicates when the difference of the GPT experiment is significant with $p <0.05$.}
\resizebox{\columnwidth}{!}{%
\begin{tabular}{lcccc}
\toprule
 & NKI-OC-VC & NKI-SpeechRT & NeuroVoz & TORGO   \\
\midrule
$n_{\text{spk}}$ & 15 & 54 & 50 & 6 \\
$n_{\text{spk-time}}$ & 26 & 138 & 50 & 6 \\
$n_{\text{sen}}$ & 92 & 23 & 13 & 55 \\
language & Dutch & Dutch & Spanish & English \\

\midrule
\multicolumn{5}{l}{\textbf{Acoustic Confounder Baselines}} \\
\midrule
Speech Rate & 0.5916 & 0.4017 & 0.0126 & 0.8422 \\
WADA SNR & -0.6546 & -0.2861 & 0.0357 & -0.2948   \\
\midrule
\multicolumn{5}{l}{\textbf{Proposed System}} \\
\midrule
n-gram ($n=5$) & -0.8617 & -0.7625 & \underline{\textbf{-0.8879}} & -0.8361 \\
\texttt{gpt-3.5-turbo} & \underline{-0.8995$\pm$0.01} & -0.8363$\pm$0.006 & -0.7095$\pm$0.006 & -0.9408$\pm$0.006 \\
\texttt{gpt-4.1} & -0.8926$\pm$0.001 & \underline{-0.8430$\pm$0.004} & -0.7627$\pm$0.013$^\dagger$ & \underline{-0.9437$\pm$0.008} \\
\midrule
\multicolumn{5}{l}{\textbf{Reference-Based Upper Bound}} \\
\midrule
WER   & \textbf{-0.9009} & \textbf{-0.8471} & -0.8248 & \textbf{-0.9507} \\
\bottomrule
\end{tabular}%
}
\label{tab:updated_results}
\end{table}

First, we confirm that neither of the acoustic confounder baselines show a consistently high correlation with all datasets as shown in Table ~\ref{tab:updated_results}. For the TORGO dataset, the speech rate shows a strong correlation ($r=0.8422$). These are followed by the moderate correlation of the NKI-OC-VC ($r=0.5916$), NKI-SpeechRT ($r=0.4017$), and the absence of a correlation for the NeuroVoz ($r=0.0126$). For WADA SNR, a strong correlation is exhibited by the NKI-OC-VC ($r=-0.6546$), with the TORGO ($r=-0.2948$) and the NKI-SpeechRT ($r=-0.2861$) showing moderate correlations, and no correlation with the NeuroVoz ($r=0.0357$).

Our proposed reference-free methods demonstrate strong performance across all four datasets, often achieving correlation levels that approach the reference-based WER upper bound. When compared to the reference-based approach, the \texttt{gpt-3.5-turbo} model was the closest for the NKI-OC-VC dataset, performing nearly identically to the reference-based WER. For the second Dutch dataset, NKI-SpeechRT, the \texttt{gpt-4.1} model proved most effective, again closely approaching the WER baseline. This trend continued for the TORGO dataset, where \texttt{gpt-4.1} also yielded the best reference-free result, showing a performance level highly competitive with the reference-based upper bound. Interestingly, for the Spanish NeuroVoz dataset, the simpler n-gram-based method achieved the best overall result, even outperforming the reference-based WER. We believe this can be accounted for by the fact that some of the NeuroVoz sentences are extremely short (4 words), and therefore do not benefit from an LLM's ability to model longer-range dependencies.

\subsection{RQ2: Language model choice}
This question assesses how the choice of language model (n-gram vs. LLMs) affects performance. The results clearly show that the more advanced LLMs generally lead to better intelligibility estimation. Moving from a 5-gram model to \texttt{gpt-3.5-turbo} or \texttt{gpt-4.1} yields a performance boost on three out of the four datasets. For instance, on NKI-SpeechRT, the correlation improves from $r=-0.7625$ (n-gram) to $r=-0.8430$ (\texttt{gpt-4.1}). While the LLMs were generally better than the n-gram model, the only statistically significant performance difference between the two GPT models was observed on the NeuroVoz dataset. This suggests that while larger language models are beneficial, the degree of improvement can be dataset-dependent.

\begin{table}[h!]
\centering
\caption{Examples of greedy, LLM corrected, and reference  transcriptions showing loss of intelligibility due to lack of precision with plosives. \textcolor{red}{\textbf{Red boldface}} highlights the differences. }
\label{tab:examples}
\resizebox{\columnwidth}{!}{
\begin{tabular}{ll}
\toprule
$W_{\text{greedy}}$ & $W_{\text{LLM}}$ \\
\midrule
de tortelduif zonk klagelijk in de oude beu\textcolor{red}{\textbf{l}} & de tortelduif zonk klagelijk in de oude beu\textcolor{red}{\textbf{k}} \\
Correct & de tortelduif zonk klagelijk in de oude beuk\\
\midrule

zij hadden \textcolor{red}{\textbf{in}} een knooi\textcolor{red}{\textbf{t}} in de hand &  zij hadden \textcolor{red}{\textbf{.}} een knoo\textcolor{red}{\textbf{p}} in de hand \\
Correct & zij hadden \textcolor{red}{\textbf{ieder}} een \textcolor{red}{\textbf{cadeautje}} in de hand \\
\bottomrule
\end{tabular}
}
\end{table}

\begin{table}[h!]
\centering \caption{Word Error Rate (WER) across all speakers by dataset. $r_{W_{\text{LLM}}}$ is the correlation between the intelligibility scores and $W_{\text{LLM}}$ WER for each speaker.}
\label{tab:wer_comparison}
\resizebox{\columnwidth}{!}{
\begin{tabular}{lcccc}
\toprule
\textbf{Metric} & \textbf{NKI-OC-VC} & \textbf{NKI-SpeechRT} & \textbf{NeuroVoz} & \textbf{TORGO} \\ \midrule $W_{\text{greedy}}$ WER & 50.04\% & 41.62\% & 27.35\% & 57.81\% \\
$W_{\text{LLM}}$ WER & 44.73\% & 33.79 \% & 20.13\% & 54.70\% \\
\midrule
$r_{W_{\text{LLM}}}$ &
-0.8969 & -0.8024 & -0.8634 & -0.9394 \\
\bottomrule
\end{tabular}
}
\end{table}

\subsection{RQ3: Explainability}

Our method's explainability lies in the difference between the greedy and the corrected transcriptions. Table~\ref{tab:examples} illustrates this process. In the first example, the model corrects a word-final plosive error. This is a common articulatory precision problem and provides a clinically meaningful reason for the utterance's reduced intelligibility. The second example shows a poorer explanation; while the resulting error score from the sentence is appropriate, the specific edits do not cleanly isolate the underlying problematic sound (/tj/). These examples show that our method is not only explainable but certain examples correspond to meaningful articulation problems.

The accuracy of the transcriptions is summarized in
Table~\ref{tab:wer_comparison}. The findings consistently indicate that the $W_{\text{LLM}}$ WER is lower than the $W_{\text{greedy}}$ WER across all evaluated datasets, ranging from 3.11\% absolute improvement for the TORGO to 7.83\% for the NKI-SpeechRT. However, we also find that the $W_{\text{LLM}}$ still has a strong correlation with the intelligibility scores.

These results show that the LLM improves reference quality but it does not always reconstruct it perfectly. This suggests the mechanism for estimating intelligibility is more nuanced than outlined in the introduction. For less severe speakers, the proposed mechanism holds true. However, for more severe speakers, the key mechanism is the disagreement between the greedy and the LLM transcription. Therefore, another interpretation is that our method measures uncertainty by measuring the inconsistency between the two ASR "listeners". Nevertheless, our proposed approach makes the estimation process more transparent than existing reference-free methods.

\section{Conclusion}
In this work, we introduced the ASR Inconsistency Score, a novel, reference-free, and explainable method for evaluating the intelligibility of pathological speech. Our approach quantifies intelligibility by measuring the discrepancy between a direct acoustic transcription and a reference from a language model. Our experiments across four datasets in Dutch, Spanish, and English showed that this method achieves a high correlation with perceptual ratings, performing competitively with reference-based methods. We showed that using advanced LLMs generally yields the best performance, showing the importance of robust linguistic modeling in reconstructing a speaker's intent. Furthermore, while we find that the LLM-generated references are more accurate than the initial transcriptions, they are far from perfect. Future work will focus on extending this method to spontaneous speech and evaluating its effectiveness across a wider range of language families.

\section{Acknowledgment}

The data collected received ethical approval. This work is partly financed by the NWO under project number 019.232SG.011, and
partly supported by a project, JPNP25006 commissioned by NEDO.

\bibliographystyle{IEEEtran}
\bibliography{refs}

@inproceedings{kim08e_interspeech,
  title     = {Robust signal-to-noise ratio estimation based on waveform amplitude distribution analysis},
  author    = {Chanwoo Kim and Richard M. Stern},
  year      = {2008},
  booktitle = {Interspeech 2008},
  pages     = {2598--2601},
  doi       = {10.21437/Interspeech.2008-644},
  issn      = {2958-1796},
}

@article{ardilacommon,
  title={Common Voice: A Massively-Multilingual Speech Corpus},
  author={Ardila, Rosana and Branson, Megan and Davis, Kelly and Henretty, Michael and Kohler, Michael and Meyer, Josh and Morais, Reuben and Saunders, Lindsay and Tyers, Francis M and Weber, Gregor},
  journal={arXiv preprint arXiv:1912.06670},
  year={2019},
  doi={10.48550/arXiv.1912.06670
}
}

@inproceedings{clapham2012nki,
  title={NKI-CCRT Corpus-Speech Intelligibility Before and After Advanced Head and Neck Cancer Treated with Concomitant Chemoradiotherapy.},
  author={Clapham and others},
  booktitle={LREC},
  volume={4},
  pages={3350--3355},
  year={2012},
  organization={Citeseer}
}

@article{bartelds2022neural,
  title={Neural representations for modeling variation in speech},
  author={Bartelds, Martijn and de Vries, Wietse and Sanal, Faraz and Richter, Caitlin and Liberman, Mark and Wieling, Martijn},
  journal={Journal of Phonetics},
  volume={92},
  pages={101137},
  year={2022},
  publisher={Elsevier}
}

@article{liss2024operationalizing,
  title={Operationalizing clinical speech analytics: Moving from features to measures for real-world clinical impact},
  author={Liss, Julie and Berisha, Visar},
  journal={Journal of Speech, Language, and Hearing Research},
  volume={67},
  number={11},
  pages={4226--4232},
  year={2024},
  publisher={American Speech-Language-Hearing Association}
}

@book{bomans2013vijvervrouw,
  title={De vijvervrouw en andere sprookjes},
  author={Bomans, Godfried},
  year={2013},
  publisher={Boekerij}
}

@inproceedings{mallela2020voice,
  title={Voice based classification of patients with amyotrophic lateral sclerosis, Parkinson’s disease and healthy controls with CNN-LSTM using transfer learning},
  author={Mallela, Jhansi and Illa, Aravind and Udupa, Sathvik and Belur, Yamini and Atchayaram, Nalini and Yadav, Ravi and Reddy, Pradeep and Gope, Dipanjan and Ghosh, Prasanta Kumar and others},
  booktitle={ICASSP 2020-2020 IEEE International Conference on Acoustics, Speech and Signal Processing (ICASSP)},
  pages={6784--6788},
  year={2020},
  organization={IEEE}
}

@article{de1997test,
  title={Test-retest study of the GRBAS scale: influence of experience and professional background on perceptual rating of voice quality},
  author={De Bodt, Marc S and Wuyts, Floris L and Van de Heyning, Paul H and Croux, Christophe},
  journal={Journal of Voice},
  volume={11},
  number={1},
  pages={74--80},
  year={1997},
  publisher={Elsevier}
}

@article{landa2014association,
  title={Association between objective measurement of the speech intelligibility of young people with dysarthria and listener ratings of ease of understanding},
  author={Landa, Sophie and Pennington, Lindsay and Miller, Nick and Robson, Sheila and Thompson, Vicki and Steen, Nick},
  journal={International Journal of Speech-Language Pathology},
  volume={16},
  number={4},
  pages={408--416},
  year={2014},
  publisher={Taylor \& Francis}
}

@inproceedings{janbakhshi2019pathological,
  title={Pathological speech intelligibility assessment based on the short-time objective intelligibility measure},
  author={Janbakhshi, Parvaneh and Kodrasi, Ina and Bourlard, Herv{\'e}},
  booktitle={ICASSP 2019-2019 IEEE International Conference on Acoustics, Speech and Signal Processing (ICASSP)},
  pages={6405--6409},
  year={2019},
  organization={IEEE}
}

@inproceedings{schu2023using,
  title={On using the UA-Speech and TORGO databases to validate automatic dysarthric speech classification approaches},
  author={Schu, Guilherme and Janbakhshi, Parvaneh and Kodrasi, Ina},
  booktitle={ICASSP 2023-2023 IEEE International Conference on Acoustics, Speech and Signal Processing (ICASSP)},
  pages={1--5},
  year={2023},
  organization={IEEE}
}

@inproceedings{liu24f_interspeech,
  title     = {Clever Hans Effect Found in Automatic Detection of Alzheimer's Disease through Speech},
  author    = {Yin-Long Liu and Rui Feng and Jia-Hong Yuan and Zhen-Hua Ling},
  year      = {2024},
  booktitle = {Interspeech 2024},
  pages     = {2435--2439},
  doi       = {10.21437/Interspeech.2024-1018},
  issn      = {2958-1796},
}

@inproceedings{son01_eurospeech,
  author={Rob J. J. H. van Son and others},
  title={{The IFA corpus: a phonemically segmented dutch "open source" speech database}},
  year=2001,
  booktitle={Proc. 7th European Conference on Speech Communication and Technology (Eurospeech 2001)},
  pages={2051--2054},
  doi={10.21437/Eurospeech.2001-484}
}

@article{halpern2023automatic,
  title={Automatic evaluation of spontaneous oral cancer speech using ratings from naive listeners},
  author={Halpern, Bence Mark and Feng, Siyuan and van Son, Rob and van den Brekel, Michiel and Scharenborg, Odette},
  journal={Speech Communication},
  volume={149},
  pages={84--97},
  year={2023},
  publisher={Elsevier}
}

@inproceedings{speech_severity_evaluation_2024,
  title = {Reference-free automatic speech severity evaluation using acoustic unit language modelling},
  author = {Bence Mark Halpern and Tomoki Toda},
  year = {2024},
  copyright = {rightsretained},
  booktitle = {Proceedings of the ACM Multimedia Asia Workshops (MMASIA Workshops '24)},
  address = {Auckland, New Zealand},
  month = {December 3--6},
  publisher = {Association for Computing Machinery},
  doi = {10.1145/3700410.3702114},
  isbn = {979-8-4007-1314-9/24/12}
}

@inproceedings{halpern2023improving,
  title={Improving severity preservation of healthy-to-pathological voice conversion with global style tokens},
  author={Halpern, Bence Mark and Huang, Wen-Chin and Violeta, Lester Phillip and van Son, RJJH and Toda, Tomoki},
  booktitle={2023 IEEE Automatic Speech Recognition and Understanding Workshop (ASRU)},
  pages={1--7},
  year={2023},
  organization={IEEE}
}

@article{maier2009automatic,
  title={Automatic speech recognition systems for the evaluation of voice and speech disorders in head and neck cancer},
  author={Maier, Andreas and Haderlein, Tino and Stelzle, Florian and N{\"o}th, Elmar and Nkenke, Emeka and Rosanowski, Frank and Sch{\"u}tzenberger, Anne and Schuster, Maria},
  journal={EURASIP Journal on Audio, Speech, and Music Processing},
  volume={2010},
  pages={1--7},
  year={2009},
  publisher={Springer}
}

@article{mendes2024neurovoz,
  title={NeuroVoz: a Castillian Spanish corpus of parkinsonian speech},
  author={Mendes-Laureano and others},
  journal={Scientific Data},
  volume={11},
  number={1},
  pages={1367},
  year={2024},
  publisher={Nature Publishing Group UK London}
}

@article{rudzicz2012torgo,
  title={The TORGO database of acoustic and articulatory speech from speakers with dysarthria},
  author={Rudzicz, Frank and Namasivayam, Aravind Kumar and Wolff, Talya},
  journal={Language resources and evaluation},
  volume={46},
  pages={523--541},
  year={2012},
  publisher={Springer}
}

@article{ozbolt2022things,
  title={Things to consider when automatically detecting Parkinson’s disease using the phonation of sustained vowels: analysis of methodological issues},
  author={Ozbolt, Alex S and Moro-Velazquez, Laureano and Lina, Ioan and Butala, Ankur A and Dehak, Najim},
  journal={Applied Sciences},
  volume={12},
  number={3},
  pages={991},
  year={2022},
  publisher={Multidisciplinary Digital Publishing Institute}
}

@article{keidser2020quest,
  title={The quest for ecological validity in hearing science: What it is, why it matters, and how to advance it},
  author={Keidser, Gitte and others},
  journal={Ear and hearing},
  volume={41},
  pages={5S--19S},
  year={2020},
  publisher={LWW}
}

@article{joshy2023dysarthria,
  title={Dysarthria severity assessment using squeeze-and-excitation networks},
  author={Joshy, Amlu Anna and Rajan, Rajeev},
  journal={Biomedical Signal Processing and Control},
  volume={82},
  pages={104606},
  year={2023},
  publisher={Elsevier}
}

@INPROCEEDINGS{10094857,
  author={Javanmardi, Farhad and others},
  booktitle={ICASSP 2023 - 2023 IEEE International Conference on Acoustics, Speech and Signal Processing (ICASSP)}, 
  title={Wav2vec-Based Detection and Severity Level Classification of Dysarthria From Speech}, 
  year={2023},
  volume={},
  number={},
  pages={1-5},
  doi={10.1109/ICASSP49357.2023.10094857}}

@article{enderby1980frenchay,
  title={Frenchay dysarthria assessment},
  author={Enderby, Pamela},
  journal={British Journal of Disorders of Communication},
  volume={15},
  number={3},
  pages={165--173},
  year={1980},
  publisher={Taylor \& Francis}
}

@article{tripathi2021automatic,
  title={Automatic speaker independent dysarthric speech intelligibility assessment system},
  author={Tripathi, Ayush and Bhosale, Swapnil and Kopparapu, Sunil Kumar},
  journal={Computer Speech \& Language},
  volume={69},
  pages={101213},
  year={2021},
  publisher={Elsevier}
}

@article{gupta2021residual,
  title={Residual neural network precisely quantifies dysarthria severity-level based on short-duration speech segments},
  author={Gupta, Siddhant and others},
  journal={Neural Networks},
  volume={139},
  pages={105--117},
  year={2021},
  publisher={Elsevier}
}

\end{document}